\documentclass[12pt,a4paper]{article} 
\usepackage{graphicx}
\newcommand{\vs}{\vspace{-0.25cm}}
\begin{document} 

\begin{center}
{\Large{\bf Quartic isospin asymmetry energy of nuclear matter\\ from 
chiral pion-nucleon dynamics}\footnote{This work has been supported in part by 
DFG and NSFC (CRC 110).}}
\medskip

N. Kaiser\\

\smallskip

{\small Physik Department T39, Technische Universit\"{a}t M\"{u}nchen,
    D-85747 Garching, Germany \\
\smallskip

{\it email: nkaiser@ph.tum.de}}
\end{center}

\medskip

\begin{abstract}
Based on a chiral approach to nuclear matter, we calculate the quartic term 
in the expansion of the equation of state of isospin-asymmetric nuclear matter.
The contributions to the quartic isospin asymmetry energy $A_4(k_f)$ arising 
from $1\pi$-exchange and chiral $2\pi$-exchange in nuclear matter are 
calculated analytically together with three-body terms involving virtual
$\Delta(1232)$-isobars. From these interaction terms one obtains at 
saturation density $\rho_0 = 0.16\,$fm$^{-3}$ the value $A_4(k_{f0})= 
1.5\,$MeV, more than three times as large as the kinetic energy part. 
Moreover, iterated $1\pi$-exchange exhibits components for which the fourth 
derivative with the respect to the isospin asymmetry parameter $\delta$ becomes 
singular at $\delta =0$. The genuine presence of a non-analytical term 
$\delta^4 \ln|\delta|$ in the expansion of the energy per particle of 
isospin-asymmetric nuclear matter is demonstrated by evaluating a s-wave 
contact interaction at second order.     
\end{abstract}



\section{Introduction and summary}
The determination of the equation of state of isospin-asymmetric nuclear 
matter has been a longstanding goal shared by both nuclear physics and 
astrophysics \cite{topical}. Usually one assumes a parabolic form for the 
energy per nucleon at zero temperature, $\bar E_{\rm as}(\rho_p,\rho_n)= \bar 
E(\rho)+ A_2(\rho)\,\delta^2+{\cal O}(\delta^4)$, where $\rho = \rho_p+\rho_n$ 
is the total nucleon density and $\delta= (\rho_n-\rho_p)/\rho$ is the isospin 
asymmetry related to unequal proton and neutron densities $\rho_p \ne 
\rho_n$.  The validity of the quadratic approximation has been verified with 
good numerical accuracy from isospin-symmetric nuclear matter $(\delta =0)$ up
to pure neutron matter $(\delta =1)$ by most of the existing nuclear many-body 
theories using various interactions \cite{quadrat}. Nonetheless, it has been 
shown consistently in numerous studies \cite{astro} that for some properties of 
neutron stars, such as the proton fraction at beta-equilibrium, the core-crust
transition density and the critical density for the direct URCA process to 
occur, even a very small quartic isospin asymmetry energy $A_4(\rho)$ 
(multiplied with $\delta^4$ in the expansion of the energy per nucleon) can 
make a big difference.    

Given the fact that all the available numerical solutions of the nuclear 
many-body problem confirm the validity of the quadratic approximation, the 
quartic isospin asymmetry $A_4(\rho)$ should be rather small. However, in the 
recent work by Cai and Li \cite{CaiLi}, which employs an empirically 
constrained isospin-dependent single-nucleon momentum-distribution and the 
equation of state of pure neutron matter near the unitary limit, a significant 
quartic isospin asymmetry energy of  $A_4(\rho_0)= (7.2\pm 2.5)\,$MeV has been 
found. This value  amounts to about 16 times the free Fermi gas prediction, see 
eq.(2). On the other hand, the calculations by the Darmstadt group 
\cite{drischl} based on chiral low-momentum interactions and many-body 
perturbation theory lead to a small value of $A_4(\rho_0)=(1.0\pm 0.2)\,$MeV. 

The purpose of the present paper it to give a prediction for the 
density-dependent quartic isospin asymmetry energy $A_4(k_f)$ in the chiral 
approach to nuclear matter developed in refs.\cite{nucmat1,nucmat2}. In this 
approach the long- and medium-range NN-interactions arising from multi-pion 
exchange are treated explicitly and few parameters encoding the relevant 
short-distance dynamics are adjusted to bulk properties of nuclear matter.
Single-particle potentials \cite{nucmat2}, quasi-particle interactions 
\cite{landau}, the thermodynamic behavior of nuclear matter at finite 
temperatures \cite{thermo} and the density-dependence of the in-medium quark 
condensate \cite{condens} follow then as predictions in that framework (see 
also the recent review article \cite{ourreview}).

The present paper is organized as follows. In section\,2, analytical 
expressions are given for the contributions to the quartic isospin asymmetry 
energy $A_4(k_f)$ as they arise from $1\pi$-exchange 
and chiral $2\pi$-exchange. The three-nucleon interaction generated by 
$2\pi$-exchange and excitation of a virtual $\Delta(1232)$-isobar is 
considered as well. These interaction contributions lead at saturation density 
$\rho_0 = 0.16\,$fm$^{-3}$ (or $k_{f0} =263\,$MeV) to the (small) value 
$A_4(k_{f0})= 1.5\, $MeV, which amounts to about three times the kinetic 
energy part. Moreover, in the course of the calculation one encounters 
components of the second-order $1\pi$-exchange whose representation of the 
fourth derivative with the respect to $\delta$ at $\delta=0$ is singular. In 
section\,3, the generic presence of a non-analytical term 
$\delta^4 \ln|\delta|$ in the expansion of the energy per particle of 
isospin-asymmetric nuclear matter is demonstrated by calculating in closed 
form the second-order contribution from a s-wave contact-interaction. Clearly, 
after having established its existence, the non-analytical term $\delta^4 
\ln|\delta|$ should be included in future fits of the equation of state of 
(zero-temperature) isospin-asymmetric nuclear matter.  

\section{One-pion and two-pion exchange contributions} 
In this section we collect the expressions for the quartic isospin asymmetry 
$A_4(k_f)$ as they arise from one-pion and two-pion exchange diagrams following
refs.\cite{nucmat1,nucmat2}. Isospin-asymmetric (spin-saturated) nuclear 
matter is characterized by different proton and neutron Fermi momenta,
$k_{p,n} =k_f(1\mp \delta)^{1/3}$. Expanding the energy per particle at fixed 
nucleon density $\rho = 2k_f^3/3\pi^2$ in the isospin asymmetry parameter 
$\delta$ up to fourth order gives:
\begin{equation} \bar E_{\rm as}(k_p,k_n) = \bar E(k_f) + \delta^2  A_2(k_f) +  
\delta^4 A_4(k_f) + {\cal O}(\delta^6) \,, \end{equation}
with $A_2(k_f)$ the (usual) quadratic isospin asymmetry energy. 
We view the density-dependent expansion coefficients $\bar E(k_f), A_2(k_f)$ and
$A_4(k_f)$ as functions of the Fermi momentum $k_f$, since they emerge in this 
form directly from the calculation. The first contribution to $A_4(k_f)$ comes 
from the relativistically improved kinetic energy $T_{\rm kin}(p)=p^2/2M-
p^4/8M^3$, and it reads:
\begin{equation} A_4(k_f)^{\rm (kin)} = {k_f^2\over 162M}\bigg(1+{k_f^2\over 4M^2}
\bigg)\,, \end{equation}
with $M=939\,$MeV the average nucleon mass. The corresponding value at nuclear 
matter saturation density $\rho_0 = 0.16\,$fm$^{-3}$ (or $k_{f0} =263\,$MeV) is 
$A_4(k_{f0})^{\rm (kin)} = 0.464\,$MeV. 

For the treatment 
of two-body interactions that depend on the momentum transfer $|\vec p_1\!-\!
\vec p_2|$ the following expansion formulas for integrals over two Fermi 
spheres are most helpful:
\begin{eqnarray}&& \int\!{d^3p_1d^3p_2\over (2\pi)^6} \, F(|\vec p_1\!-\!\vec 
p_2|) \Big[\theta(k_p-|\vec  p_1|)\, \theta(k_p-|\vec p_2|)+ \theta(k_n-|\vec 
p_1|)\, \theta (k_n-|\vec p_2|)\Big]\nonumber \\ && ={2k_f^6 \over 3\pi^4}
\int_0^1\!\!dz \bigg\{\bigg[z^2(1-z)^2(2+z) +{\delta^2z^3 \over 3}\bigg] 
F(2z k_f) +{\delta^4 k_f\over 162} \Big[ F'(2k_f) - 7 z^4 F'(2z k_f) \Big] 
\bigg\}  \,,\end{eqnarray}
\begin{eqnarray}&& \int\!{d^3p_1d^3p_2\over(2\pi)^6}\,F(|\vec p_1\!-\!\vec p_2|) 
\,\theta(k_p-|\vec  p_1|)\, \theta(k_n-|\vec p_2|) \nonumber \\ && = {k_f^6 
\over 3\pi^4}\int_0^1\!\!dz \bigg\{\bigg[z^2(1-z)^2(2+z) +{\delta^2z \over 3}
(z^2-1)\bigg] F(2z k_f) +{\delta^4 k_f\over 162}(8z^2-1-7z^4) F'(2z k_f) 
\bigg\}  \,.\end{eqnarray}
The $z$-dependent weighting functions at order $\delta^2$ and $\delta^4$ have 
been obtained by applying several partial integrations. The contribution of the 
$1\pi$-exchange Fock diagram to the quartic isospin asymmetry energy reads:
\begin{eqnarray} A_4(k_f)^{\rm (1\pi)} &=&  {g_A^2 m_\pi^3\over (36\pi f_\pi)^2}
\bigg\{\bigg(4u+{21\over 8u}\bigg)\ln(1+4u^2)-2u^3-{33u \over 4}-{u(9+44u^2)
\over 4(1+4u^2)^2} \nonumber \\ && +{m_\pi^2 \over M^2}\bigg[2u^5+2u^3+
{3u\over 8}-u^3\ln(1+4u^2)-{u(3+16u^2)\over 8(1+4u^2)^2}-{3u^2 \over 2}
\arctan 2u\bigg]\bigg\} \,,\end{eqnarray}
with the dimensionless variable $u = k_f/m_\pi$. The second line in eq.(5) 
gives the relativistic $1/M^2$-correction. The occurring physical 
parameters are: nucleon axial-vector coupling constant $g_A=1.3$, 
(neutral) pion mass $m_\pi = 135\,$MeV,  and pion decay constant 
$f_\pi = 92.4\,$MeV. 

Next in the chiral expansion comes the iterated (second-order) 
$1\pi$-exchange. With two medium insertions ${1\over 2}(1+\tau_3)\theta
(k_p-|\vec p_i|)+{1\over 2}(1-\tau_3)\theta(k_n-|\vec p_i|)$ one gets a 
Hartree contribution of the form:
\begin{equation} A_4(k_f)^{\rm (H2)} = {g_A^4 M m_\pi^4 \over (24\pi)^3f_\pi^4}
\bigg\{10u^3-{61u \over 2}+{200u^2+49 \over 6u}\ln(1+4u^2)-{u(13+60u^2) \over 
6(1+4u^2)^2}-{128u^2 \over 3}\arctan 2u\bigg\}\,, \end{equation}
and the corresponding Fock exchange-term reads:
\begin{eqnarray} A_4(k_f)^{\rm (F2)} &=&{g_A^4 M m_\pi^4 \over (12\pi)^3f_\pi^4}
\bigg\{{u \over 8}-{u^3 \over 3}-{u \over 12(1+2u^2)}-{u \over 24(1+u^2)} 
\nonumber \\ && +u^4 \arctan u+{u^2(2+11u^2+16u^4)\over 6(1+2u^2)^2} \Big[
\arctan u-\arctan 2u\Big] \nonumber \\ && + \int_0^u \!\!dx \,{21x^2-16u^2 
\over 6u(1+2x^2)}\Big[(1+8x^2+8x^4)\arctan x-(1+4x^2)\arctan 2x \Big]\bigg\} \,.
\end{eqnarray}
Pauli-blocking effects at second order are included through diagrams with 
three (isospin-asymmetric) medium insertions \cite{nucmat1}. We consider here 
only the factorizable Fock contribution for which the energy denominator gets 
canceled by factors from the momentum-dependent $\pi N$-vertices (see 
eqs.(11,26) in ref.\cite{nucmat1}). Its contribution to the quartic isospin 
asymmetry energy can be represented as a one-parameter integral, 
$A_4(k_f)^{\rm(fac)}=g_A^4M m_\pi^4(12\pi f_\pi)^{-4}\int_0^u\!dx I(x,u)$, where the 
lengthy integrand $I(x,u)$ involves the function $\ln[1+(u+x)^2]-\ln[1+(u-x)^2
]$ and its square. The corresponding value at saturation density is $A_4(k_{f0}
)^{\rm(fac)}= -1.35\,$MeV, thus counterbalancing most of the Fock term $A_4(k_{f0}
)^{\rm (F2)} = 1.70\,$MeV without Pauli-blocking written in eq.(7). For the 
non-factorizable pieces the representation of the fourth derivative with 
respect to $\delta$ at $\delta=0$ includes singularities of the form 
$(u-x)^{-\nu}, \nu=1,2$. When subtracting these singular terms from the 
integrand only very small numerical values are obtained for the 
non-factorizable Hartree contribution. In the case of the quadratic 
isospin asymmetry energy $A_2(k_{f0})$ one finds that the non-factorizable 
pieces (see eqs.(24,26) in ref.\cite{nucmat1}) tend to cancel each other 
almost completely as: $(-11.6+12.0)\,$MeV. Therefore one can 
expect that the omission of the non-factorizable pieces does not change much 
the final result for the quartic isospin asymmetry energy $A_4(k_f)$. However, 
the observation that the iterated $1\pi$-exchange has components with a singular
representation of their fourth derivative with respect to $\delta$ at $\delta 
=0$, indicates that the expansion in eq.(1) becomes non-analytic beyond the 
quadratic order $\delta^2$. This feature is demonstrated in section\,3 by 
calculating in closed form the second-order contribution from a s-wave 
contact-interaction.

We continue with the contribution of the irreducible $2\pi$-exchange to the 
quartic isospin asymmetry energy. Using a twice-subtracted dispersion relation
for the $2\pi$-exchange NN-potential in momentum-space and the master formulas 
in eqs.(3,4), one obtains:  
\begin{eqnarray} A_4(k_f)^{\rm (2\pi)} &=& {1 \over 81\pi^3} \int_{2m_\pi}^\infty 
\!\! d\mu \bigg\{{\rm Im}(V_C+2\mu^2V_T)\bigg[{7\mu k_f \over 4}-{2k_f^5 \over 
3\mu^3} -{\mu k_f^3(7\mu^2+36k_f^2) \over 2(\mu^2+4k_f^2)^2}  \nonumber \\ && 
-{7\mu^3 \over 16k_f} \ln\bigg(1+{4k_f^2 \over \mu^2}\bigg) \bigg]  +{\rm Im}
(W_C+2\mu^2W_T) \bigg[{2k_f^5 \over \mu^3}+{k_f^3 \over \mu}+{21\mu k_f \over 4}
 \nonumber \\ && -{\mu k_f^3(7\mu^2+36k_f^2) \over 2(\mu^2+4k_f^2)^2}
-{\mu \over 16k_f}(21\mu^2+32k_f^2) \ln\bigg(1+{4k_f^2 \over \mu^2}\bigg) 
\bigg]\bigg\}\,,  \end{eqnarray}
where Im$V_{C,T}$ and Im$W_{C,T}$ are the spectral functions of the isoscalar 
and isovector central and tensor NN-amplitudes, respectively. These imaginary 
parts are composed of the functions $\sqrt{\mu^2-4m_\pi^2}$ and $\arctan(
\sqrt{\mu^2-4m_\pi^2}/2\Delta)$, with $\Delta = 293\,$MeV the delta-nucleon 
mass splitting. Note that due to the implemented subtractions the 
$k_f$-expansion of $A_4(k_f)^{\rm (2\pi)}$ in eq.(8) starts with the power $k_f^7$.
A short-distance contribution proportional to $k_f^5$ is supplemented by the 
subtraction constants: 
\begin{equation} A_4(k_f)^{\rm (sc)} =  {10k_f^5 \over (3M)^4}\bigg({2B_5\over 3}
-B_{n,5}\bigg)\,,\end{equation} 
with the parameters $B_5 = 0$ and $B_{n,5} = -3.58$ 
adjusted in ref.\cite{nucmat2} to the empirical nuclear matter saturation 
point and quadratic isospin asymmetry energy $A_2(k_{f0})^{\rm (emp)} = 34\,$MeV.

Finally, we consider the long-range three-nucleon interaction 
generated by $2\pi$-exchange and virtual excitation of a 
$\Delta(1232)$-isobar \cite{nucmat2}. The corresponding three-body Hartree  
contribution reads: 
\begin{equation}A_4(k_f)^{\rm (\Delta)}={g_A^4 m_\pi^6u^2\over \Delta(6\pi 
f_\pi)^4} \bigg\{\bigg({16u^2 \over 3}+{21 \over 4}\bigg) \ln(1+4u^2)-{4u^4 
\over 3} -{41u^2 \over 3}-{2u^2(11+99u^2+236u^4) \over 3(1+4u^2)^3} \bigg\}\,,
\end{equation} 
while the associated three-body Fock term can be represented as 
$g_A^4 m_\pi^6 (12\pi f_\pi)^{-4} \Delta^{-1}\int_0^u\!dx J(x,u)$, where the lengthy
integrand $J(x,u)$ involves the functions $\arctan(u+x)+ \arctan(u-x)$ 
and $\ln[1+(u+x)^2]-\ln[1+(u-x)^2]$. Note that the three-body contact-term 
proportional to $\zeta$ introduced additionally in ref.\cite{nucmat2} does not 
contribute to the quartic isospin asymmetry energy $A_4(k_f)$.
\begin{figure}
\begin{center}
\includegraphics[scale=0.48,clip]{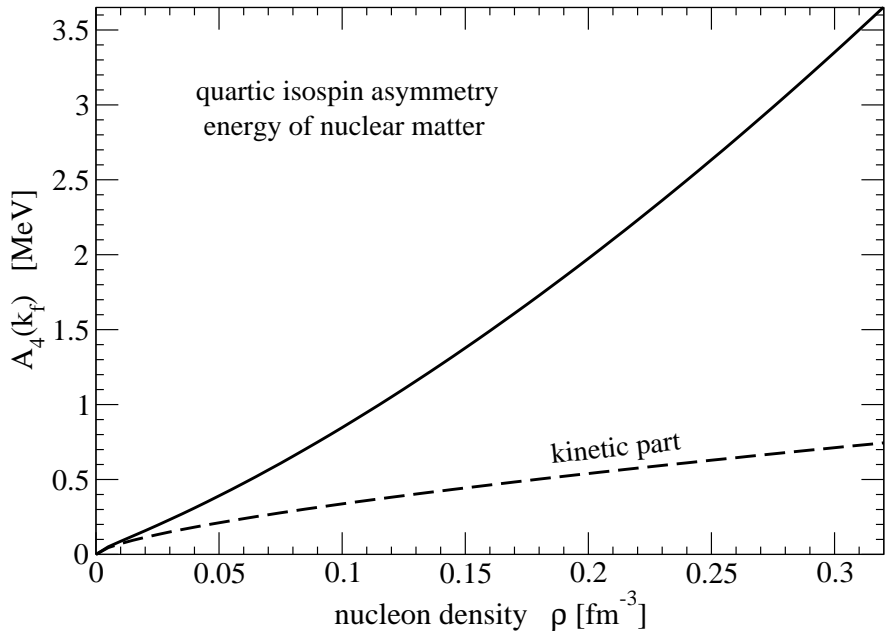}
\end{center}
\vspace{-.6cm}
\caption{Quartic isospin asymmetry $A_4(k_f)$ as a function of the 
nucleon density $\rho=2k_f^3/3\pi^2$.}
\end{figure}

Summing up all the calculated contributions, one obtains the result for the 
density-dependent quartic isospin asymmetry energy $A_4(k_f)$ of nuclear matter 
as shown in Fig.\,1 in the density region $0<\rho<2\rho_0=0.32\,$fm$^{-3}$. 
The predicted value at saturation density $\rho_0 = 0.16\,$fm$^{-3}$ is 
$A_4(k_{f0})=  1.49\,$MeV. It amounts to $3.2$ times  
the free Fermi-gas part $A_4(k_{f0})^{\rm (kin)}=  0.464\,$MeV. Note that 
interaction contributions to  $A_4(k_f)$ start (at least) with the power 
$k_f^5$. The density-dependence of the full line in Fig.\,1 is to a good 
approximation $\rho^{5/4}$. Actually, it should be noted that the present 
calculation of the quartic isospin asymmetry energy $A_4(k_f)$ is performed in 
a framework where empirical constraints from bulk properties of nuclear matter 
are satisfied. This does not apply to the recent work in ref.\cite{CaiLi}, 
where a large value of $A_4(k_{f0})= (7.2\pm 2.5)\,$MeV has been found.     

\section{S-wave contact interaction to second order}
The analysis of the Pauli-blocking corrections to the second-order (iterated) 
$1\pi$-exchange has indicated that non-analytical terms may occur in the 
$\delta$-expansion of the energy per particle of isospin-asymmetric nuclear 
matter beyond the quadratic order. In the extreme case there could be a 
cubic term $|\delta|^3$, which is after all even under the exchange of protons 
and neutrons: $\delta\to -\delta$. In order to clarify the situation, we 
consider a s-wave contact interaction:
\begin{equation} V_{\rm ct} = {\pi \over M}\Big[ a_s+3a_t +(a_t-a_s)\,\vec 
\sigma_1 \cdot \vec \sigma_2\Big]\,,\end{equation} 
and examine it in second-order many-body perturbation theory. For this simple 
interaction, the occurring integrals over three Fermi spheres with (at most) 
two different radii $k_p, k_n$ can be solved in closed analytical form. The 
pertinent function to express the result in the isospin-asymmetric 
configuration of interest is:
\begin{eqnarray} && 35\!\int_0^1 \!dz\,(z-z^4) \bigg\{2x z+(x^2-z^2)\ln{x+z 
\over |x-z|}\bigg\} \nonumber \\ && = {x\over 2}(15+33x^2-4x^4) +{1\over 4} 
(42x^2-15-35x^4)\ln{x+1 \over|x-1|}+2x^7\ln{x^2\over |x^2-1|}\,,\end{eqnarray} 
where the variable $x>0$ is set to a ratio of Fermi momenta, 
$[(1+\delta)/(1-\delta)]^{\pm 1/3}$ or $1$. Note that the function defined in 
eq.(12) has  at $x=1$ the value $22-4\ln 2$. Combining the second-order Hartree 
and Fock diagrams generated by $V_{\rm ct}$ according to their spin- and 
isospin-factors and performing the expansion in powers of $\delta$, one obtains
the following result for the energy per particle: 
\begin{eqnarray} \bar E_{\rm as}(k_p,k_n)^{\rm (2nd)} &\!\!\!=\!\!\!& {k_f^4 \over 
5\pi^2 M} \bigg\{{3 \over 7} (a_s^2+a_t^2) (11-2\ln 2) +{4 \delta^2 \over 3} 
\Big[ a_s^2(3-\ln 2)-a_t^2(2+\ln 2)\Big] \nonumber \\ && +{\delta^4 \over 81} 
\bigg[a_s^2 \bigg( 10 \ln {|\delta|\over 3} + 2\ln 2 -{41 \over 6}\bigg) 
+a_t^2 \bigg(30 \ln {|\delta|\over 3}+2\ln 2 +{3\over 2}\bigg) \bigg]
+{\cal O}(\delta^6) \bigg\}\,.\end{eqnarray} 
The crucial and novel feature which becomes evident from this expression is the 
presence of the non-analytical logarithmic term $\delta^4 \ln(|\delta|/3)$. 
Interestingly, the corresponding coefficient is three times as large in the 
spin-triplet channel as in the spin-singlet channel. For comparison the 
first-order contribution of the s-wave contact interaction $V_{\rm ct}$ reads, 
$\bar E_{\rm as}(k_p,k_n)^{\rm (1st)}=  k_f^3\big[-a_s-a_t+\delta^2(a_t-a_s/3)\big]
/2\pi M$, without any higher powers of $\delta$. Note that the sign-convention 
for the scattering lengths $a_{s,t}$ is chosen here such that positive values 
correspond to attraction. As a check we have rederived the same results at 
first and second order by using the alternative (and equivalent) form of the 
s-wave contact interaction $V'_{\rm ct}=\pi [ 3a_s+a_t +(a_s-a_t)\,\vec \tau_1
\cdot \vec\tau_2]/M$.

In Fig.\,2 the dependence of the second-order energy per particle 
$\bar E_{\rm as}(k_p,k_n)^{\rm (2nd)}$ on the isospin asymmetry parameter $\delta $ 
is shown for three different choices of the s-wave scattering lengths: $a_s=
a_t$, $a_t=0$ and $a_s=0$. In each case the full line shows the exact result and
the (nearby) dashed line gives the expansion in powers of $\delta$ truncated 
at fourth order according to eq.(13). One observes that these expansions  
reproduce the full $\delta$-dependence very well over the whole range $-1\leq 
\delta \leq 1$. Note also that the prefactor 
$k_f^4 a_{s,t}^2/5\pi^2 M $ of dimension energy has been scaled out in Fig.\,2. 
\begin{figure}
\begin{center}
\includegraphics[scale=0.48,clip]{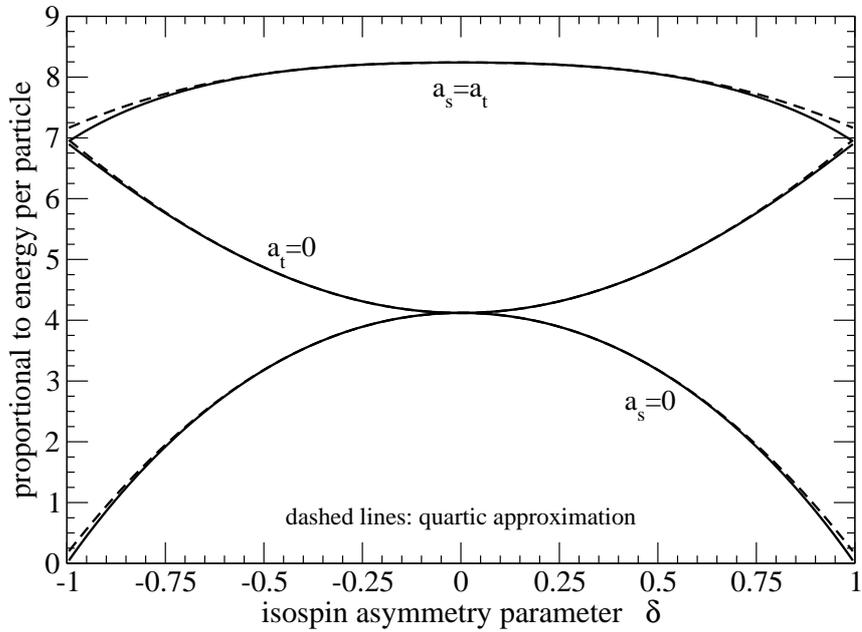}
\end{center}
\vspace{-.6cm}
\caption{Dependence of the 2nd order energy per particle $\bar E_{\rm as}(k_p,
k_n)^{\rm (2nd)}$ on the isospin asymmetry $\delta$. Three different choices for 
the scattering lengths, $a_s=a_t$, $a_t=0$ and $a_s=0$, are considered.}
\end{figure}

If one performs for the second-order energy density the fourth 
derivative with respect to $\delta$ at $\delta=0$ under the integral, then one 
encounters integrands with singularities of the form $(1-z)^{-\nu}, \nu=1,2$.  
The origin of these singularities, or in the proper treatment the 
non-analytical term $\delta^4 \ln(|\delta|/3)$, lies in the energy denominator 
of second-order diagrams. For an infinite (normal) many-fermion system the 
energy spectrum has a vanishing gap between bound states in the Fermi sea and 
excited states in the continuum.  Such a gap-less energy spectrum causes a  
singularity, respectively a non-analyticity, if small asymmetries of the Fermi 
levels of two components are analyzed with too high resolution.       

In summary, we have demonstrated that the non-analytical term $\delta^4 
\ln(|\delta|/3)$ will be generically present in calculations of isospin-asymmetric 
nuclear matter that go beyond the mean-field Hartree-Fock level. Therefore, 
a term $\delta^4 \ln|\delta|$ should be included in future fits of the 
equation of state of (zero-temperature) isospin-asymmetric nuclear matter 
and its role should be further examined.

\end{document}